\documentclass[prb,reprint,superscriptaddress]{revtex4-1}

\usepackage{graphicx}
\usepackage{dcolumn}
\usepackage{bm}
\usepackage{amsthm}
\usepackage{amsmath}
\usepackage{amssymb}
\usepackage{xcolor}

\begin{document}

\title{Evidence for the formation of nanoprecipitates with magnetically disordered regions in bulk $\mathrm{Ni}_{50}\mathrm{Mn}_{45}\mathrm{In}_{5}$ Heusler alloys}

\author{Giordano Benacchio}
\affiliation{Physics and Materials Science Research Unit, University of Luxembourg, 162A~Avenue de la Fa\"iencerie, L-1511 Luxembourg, Grand Duchy of Luxembourg}
\author{Ivan Titov}
\affiliation{Physics and Materials Science Research Unit, University of Luxembourg, 162A~Avenue de la Fa\"iencerie, L-1511 Luxembourg, Grand Duchy of Luxembourg}
\author{Artem Malyeyev}
\affiliation{Physics and Materials Science Research Unit, University of Luxembourg, 162A~Avenue de la Fa\"iencerie, L-1511 Luxembourg, Grand Duchy of Luxembourg}
\author{Inma Peral}
\affiliation{Physics and Materials Science Research Unit, University of Luxembourg, 162A~Avenue de la Fa\"iencerie, L-1511 Luxembourg, Grand Duchy of Luxembourg}
\author{Mathias Bersweiler}
\affiliation{Physics and Materials Science Research Unit, University of Luxembourg, 162A~Avenue de la Fa\"iencerie, L-1511 Luxembourg, Grand Duchy of Luxembourg}
\author{Philipp Bender}
\affiliation{Physics and Materials Science Research Unit, University of Luxembourg, 162A~Avenue de la Fa\"iencerie, L-1511 Luxembourg, Grand Duchy of Luxembourg}
\author{Denis Mettus}
\affiliation{Physik-Department, Technische Universit\"at M\"unchen, James-Franck-Stra{\ss}e, D-85748 Garching, Germany}
\author{Dirk Honecker}
\affiliation{Institut Laue-Langevin, 71~avenue des Martyrs, F-38042 Grenoble, France}
\author{Elliot Paul Gilbert}
\affiliation{Australian Centre for Neutron Scattering, ANSTO, Locked Bag 2001, Kirrawee DC, NSW 2232, Australia}
\author{Mauro Coduri}
\affiliation{European Synchrotron Radiation Facility, 6~Rue Jules Horowitz, B.P.~220, F-38043 Grenoble Cedex~9, France}
\author{Andr\'e Heinemann}
\affiliation{German Engineering Materials Science Centre (GEMS) at Heinz Maier-Leibnitz Zentrum (MLZ), Helmholtz-Zentrum Geesthacht GmbH, D-85748 Garching, Germany}
\author{Sebastian M\"uhlbauer}
\affiliation{Heinz Maier-Leibnitz Zentrum (MLZ), Technische Universit\"at M\"unchen, D-85748 Garching, Germany}
\author{Asl{\i} {\c{C}}ak{\i}r}
\affiliation{Department of Metallurgical and Materials Engineering, Mugla University, 48000 Mugla, Turkey}
\author{Mehmet Acet}
\affiliation{Faculty of Physics, Duisburg-Essen University, D-47057 Duisburg, Germany}
\author{Andreas Michels}\email{andreas.michels@uni.lu}
\affiliation{Physics and Materials Science Research Unit, University of Luxembourg, 162A~Avenue de la Fa\"iencerie, L-1511 Luxembourg, Grand Duchy of Luxembourg}

\begin{abstract}
Shell ferromagnetism is a new functional property of certain Heusler alloys which has been recently observed in $\mathrm{Ni}_{50}\mathrm{Mn}_{45}\mathrm{In}_{5}$. We report the results of a comparative study of the magnetic microstructure of bulk $\mathrm{Ni}_{50}\mathrm{Mn}_{45}\mathrm{In}_{5}$ Heusler alloys using magnetometry, synchrotron x-ray diffraction, and magnetic small-angle neutron scattering (SANS). By combining unpolarized and spin-polarized SANS (POLARIS) we demonstrate that a number of important conclusions regarding the mesoscopic spin structure can be made. In particular, the analysis of the magnetic neutron data suggests that nanoprecipitates with an effective ferromagnetic component form in an antiferromagnetic matrix on field annealing at $700 \, \mathrm{K}$. These particles represent sources of perturbation, which seem to give rise to magnetically disordered regions in the vicinity of the particle-matrix interface. Analysis of the spin-flip SANS cross section via the computation of the correlation function yields a value of $\sim 55 \, \mathrm{nm}$ for the particle size and $\sim 20 \, \mathrm{nm}$ for the size of the spin-canted region.
\end{abstract}

\maketitle

\section{Introduction}
\label{introduction}

Ni-Mn-based Heusler alloys exhibit unique physical properties such as magnetocaloric, magnetoresistance, exchange bias, and, in particular, magnetic shape-memory effects~\cite{mehmetjpcmreview2009,acet2010handbook,gutfleisch2011}. These features originate from the strongly coupled magnetic and structural degrees of freedom and occur in stoichiometric as well as in nonstoichiometric alloys. Depending on the composition, Ni-Mn-X (X:~Al, Ga, In, Sn, Sb) Heusler alloys display a martensitic phase transition from a high-temperature cubic L2$_{\mathrm{1}}$ austenite phase to a low-temperature tetragonal L1$_{\mathrm{0}}$ martensite phase. The near-stoichiometric Heusler alloys $\mathrm{Ni}_2\mathrm{Mn}\mathrm{X}$ have a ferromagnetic (FM) ground state, while the off-stoichiometric $\mathrm{Ni}_{50}\mathrm{Mn}_{50-x}\mathrm{In}_{x}$ alloys with $x < 25$ show an antiferromagnetic (AF) coupling.

A new functional property in the off-stoichiometric $\mathrm{Ni}_{50}\mathrm{Mn}_{45}\mathrm{In}_{5}$ alloy has recently been observed by~\textcite{cakir2016}. The compound, when annealed at temperatures between about $650 - 750 \, \mathrm{K}$ under the application of a magnetic field as little as $0.1 \, \mathrm{T}$, is believed to decompose into nanosized $\mathrm{Ni}_{50}\mathrm{Mn}_{25}\mathrm{In}_{25}$ precipitates which are embedded in a $\mathrm{Ni}_{50}\mathrm{Mn}_{50}$ matrix. The precipitates are FM at room temperature, whereas the matrix is AF. The spins at the interface with the NiMn matrix align with the applied field during their growth and become strongly pinned in the field direction during annealing. The remanent spin pinning persists up to temperatures of about $600 \, \mathrm{K}$, and it is estimated that an applied field of about $20 \, \mathrm{T}$ is required to reverse the spins~\cite{scheibel2017}. Consequently, the possible applications of this effect may be in nonvolatile memory. The above described core-shell structure---FM precipitates with a disordered FM shell embedded in an AF matrix---has been postulated to exist based on the outcome of integral measurement techniques~\cite{cakir2016,cakir2017}.

It is the aim of the present work to obtain microscopic information about the nanoscale structure of bulk $\mathrm{Ni}_{50}\mathrm{Mn}_{45}\mathrm{In}_{5}$ Heusler alloys by means of magnetic-field-dependent unpolarized and polarized small-angle neutron scattering (SANS). The SANS technique (see \cite{rmp2019} for a recent review) appears to be ideally suited for this endeavor, since it provides information about the variations of both the magnitude and orientation of the magnetization vector field in the bulk of a material and on a nanometer length scale ($\sim 1-100 \, \mathrm{nm}$). Magnetic SANS has previously been employed for studying the highly complex magnetic ordering and mesoscale inhomogeneity across the martensitic phase transformation in Heusler alloys. For instance, \textcite{kopitsa2003,bliznuk2004} used polarized SANS to study the effect of Si, Cr, Ni, C, and N alloying on the nuclear and magnetic homogeneity of Fe-Mn-based shape-memory alloys. \textcite{runov2001,runov2003,runov2004,runov2006} were the first to investigate the nuclear and magnetic microstructure of polycrystalline $\mathrm{Ni}_{2}\mathrm{Mn}\mathrm{Ga}$ and single crystalline $\mathrm{Ni}_{49.1}\mathrm{Mn}_{29.4}\mathrm{Ga}_{21.5}$ by means of temperature ($15 -400 \, \mathrm{K}$) and magnetic-field-dependent (up to $4.5 \, \mathrm{kOe}$) polarized SANS, neutron depolarization, and neutron diffraction; the spin dynamics was also probed via the so-called left-right asymmetry method. Their main conclusions are that all the structural phase transformations in the Ni-Mn-Ga alloy system proceed via mesoscopically inhomogeneous phases, and that the structural changes (including changes in the lattice modulation) are accompanied by changes in the spin dynamics~\cite{runov2006}. \textcite{bhatti2012,bhatti2016} scrutinized the complex magnetism of $\mathrm{Ni}_{50-x}\mathrm{Co}_x\mathrm{Mn}_{40}\mathrm{Sn}_{10}$ polycrystalline alloys with $x = 6-8$. The temperature dependence of the unpolarized zero-field SANS data (within $30-600 \, \mathrm{K}$) across $T_C = 425 \, \mathrm{K}$ and the martensitic transition at $\sim 380-390 \, \mathrm{K}$ were analyzed by a combination of a Porod, Gaussian, and Lorentzian scattering functions; these are expected to model, respectively, the scattering from large-scale structures (e.g., magnetic domains or crystal grains), nanosized spin clusters, and critical fluctuations. In agreement with conclusions drawn from magnetometry data, these authors observe the formation of nanoscopic spin clusters at low temperatures. In the context of magnetoresponsive shape-memory alloys one may also mention the SANS study of \textcite{laver2010a,laver2010b} on a magnetostrictive $\mathrm{Fe_{81}Ga_{19}}$ single crystal. Magnetic-field-dependent unpolarized and spin-polarized SANS (with and without compressive strain) revealed the existence of shape-anisotropic nanosized precipitates with a magnetization which is different than the one of the matrix phase. The role of these heterogeneities for the magnetostriction process in Fe-Ga alloys has been discussed. 

In the present study, we use SANS to evidence the formation of nanoprecipitates in field-annealed $\mathrm{Ni}_{50}\mathrm{Mn}_{45}\mathrm{In}_{5}$. The paper is organized as follows: Section~\ref{experiments} furnishes information on the sample preparation, their structural and magnetic characterization, and the details of the neutron experiments. Section~\ref{results} presents and discusses the experimental results of the magnetization, x-ray synchrotron, and neutron measurements, while Section~\ref{conclusion} summarizes the main findings of this investigation. The relevant expressions for the unpolarized and polarized SANS cross sections are summarized in the Appendix.

\section{Experimental}
\label{experiments}

$\mathrm{Ni}_{50}\mathrm{Mn}_{45}\mathrm{In}_{5}$ ingots were prepared by arc melting of high-purity elements ($99.9 \, \%$) and annealed under Ar atmosphere at $1073 \, \mathrm{K}$ in sealed quartz tubes for $5$~days for homogenization purposes. The specimens were then quenched in water at room temperature. The chemical compositions of the as-prepared samples were determined using energy-dispersive x-ray analysis (EDX). To check for sample homogeneity, EDX spectra were collected from seven different positions on the sample's surface. One $\mathrm{Ni}_{50}\mathrm{Mn}_{45}\mathrm{In}_{5}$ (nominal composition) specimen was annealed in a vibrating sample magnetometer (VSM) under an applied magnetic field of $5 \, \mathrm{T}$ at a temperature of $700 \, \mathrm{K}$ during $12 \, \mathrm{h}$, and a second one was field-annealed at $650 \, \mathrm{K}$ during $6 \, \mathrm{h}$. In the following, we refer to them as the ``$700 \, \mathrm{K}$'' and the ``$650 \, \mathrm{K}$'' samples, respectively. As reported in~\cite{cakir2016,cakir2017}, the spins at the interface between the matrix and the precipitates are assumed to align along the direction of the magnetic field applied during annealing, which we denote in the following as the (``texture'') $\mathbf{c}$-axis, and remain strongly pinned in this direction after annealing. Therefore, since the samples may be magnetically textured, the magnetization and SANS experiments were performed for two orientations of the applied magnetic field $\mathbf{H}_0$ with respect to the $\mathbf{c}$-axis. A third as-prepared sample is used as a reference, and we denote it by ``initial state''. The chemical compositions of the three samples are listed in Table~\ref{tab1}.

\begin{table}[tb!]
\caption{\label{tab1} Chemical composition (in at.$\, \%$) of the $\mathrm{Ni}_{50}\mathrm{Mn}_{45}\mathrm{In}_{5}$ samples used in this study.}
\begin{ruledtabular}
\begin{tabular}{cccc}
 & Ni & Mn & In \\ \hline
initial state & 50.6 & 44.6 & 4.8 \\
$650 \, \mathrm{K}$ & 50.3 & 44.8 & 4.9 \\
$700 \, \mathrm{K}$ & 51.2 & 43.9 & 5.0
\end{tabular}
\end{ruledtabular}
\end{table}

The neutron experiment was conducted at the D33 instrument at the Institut Laue-Langevin (ILL), Grenoble, France~\cite{dewhurst2016}. The measurements on the $650 \, \mathrm{K}$ and $700 \, \mathrm{K}$ annealed samples were made using both unpolarized and polarized incident neutrons with a mean wavelength of $\lambda = 6 \, \mathrm{\AA}$, $\Delta\lambda / \lambda = 10 \, \%$ (FWHM), and within a $q$-range of $0.04 \, \mathrm{nm}^{-1} \lesssim q \lesssim 1.5 \,\mathrm{nm}^{-1}$. The magnetic field was applied perpendicular to the incident neutron beam ($\mathbf{H}_0 \perp \mathbf{k}_0$) and either parallel to the texture axis ($\mathbf{H}_0 \parallel \mathbf{c}$) or perpendicular to it ($\mathbf{H}_0 \perp \mathbf{c}$); see Fig.~\ref{sans_setup} in the Appendix for a schematic drawing of the experimental neutron setup. For the polarized runs, the incoming neutrons were polarized by a remanent FeSi supermirror transmission polarizer ($m = 3.6$), and a radio-frequency (rf) spin flipper allowed us to reverse the initial neutron polarization. The flipping efficiency of the rf flipper was $\epsilon = 99.8 \, \%$, and the polarizer efficiency was $P = 97.6 \, \%$ at $\lambda = 6 \, \mathrm{\AA}$. The neutron experiments were performed by first applying a large positive field of $8 \, \mathrm{T}$, and then reducing the field following the magnetization curve (compare Fig.~\ref{MH1}). Further unpolarized SANS measurements under similar conditions as the ILL experiment were conducted at the QUOKKA instrument~\cite{elliot2018} at the Australian Nuclear Science and Technology Organisation (ANSTO), Lucas Heights, Australia and at SANS~1~\cite{muehlbauer2016} at the Heinz Maier-Leibnitz Zentrum, Garching, Germany. All data were collected at room temperature, except the ANSTO measurements on the initial-state sample, which were taken at $200 \, \mathrm{K}$. SANS data reduction (correction for background scattering, sample transmission, detector efficiency, spin leakage) was carried out using the GRASP software package~\cite{graspurl}.

High-energy synchrotron X-ray diffraction measurements ($\lambda = 0.190693 \, \mathrm{\AA}$, $E = 65 \, \mathrm{keV}$) were carried out in transmission mode at the beamline ID22 at the European Synchrotron Radiation Facility (ESRF), Grenoble, France~\cite{esrfid22}. The high x-ray energy was chosen to probe the sample in the same scattering geometry as in the SANS experiment. This allows us to have a direct correspondence between the structural and magnetic information obtained by the two techniques. Magnetization measurements were conducted using a $14 \, \mathrm{T}$ vibrating sample magnetometer.

\section{Results and discussion}
\label{results}

\subsection{Magnetometry}

\begin{figure}[b!]
\centering
\includegraphics[width=1.0\columnwidth]{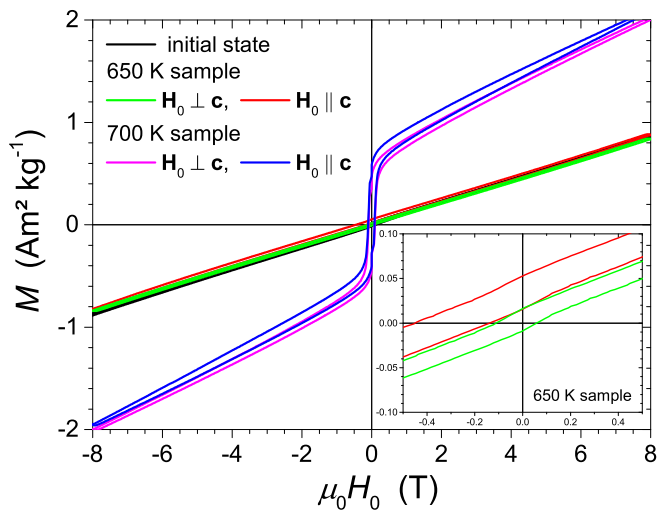}
\caption{Room-temperature magnetization curves of the $650 \, \mathrm{K}$ and $700 \, \mathrm{K}$ samples for two directions of the applied magnetic field $\mathbf{H}_0$ with respect to the texture axis $\mathbf{c}$ ($\mathbf{H}_0 \perp \mathbf{c}$ and $\mathbf{H}_0 \parallel \mathbf{c}$). The magnetization curve of the initial-state sample was recorded at $200 \, \mathrm{K}$. Inset: Zoom into the low field part of the $M(H_0)$ curves of the $650 \, \mathrm{K}$ sample.}
\label{MH1}
\end{figure}

\begin{figure*}[tb!]
\centering
\includegraphics[width=1.05\columnwidth]{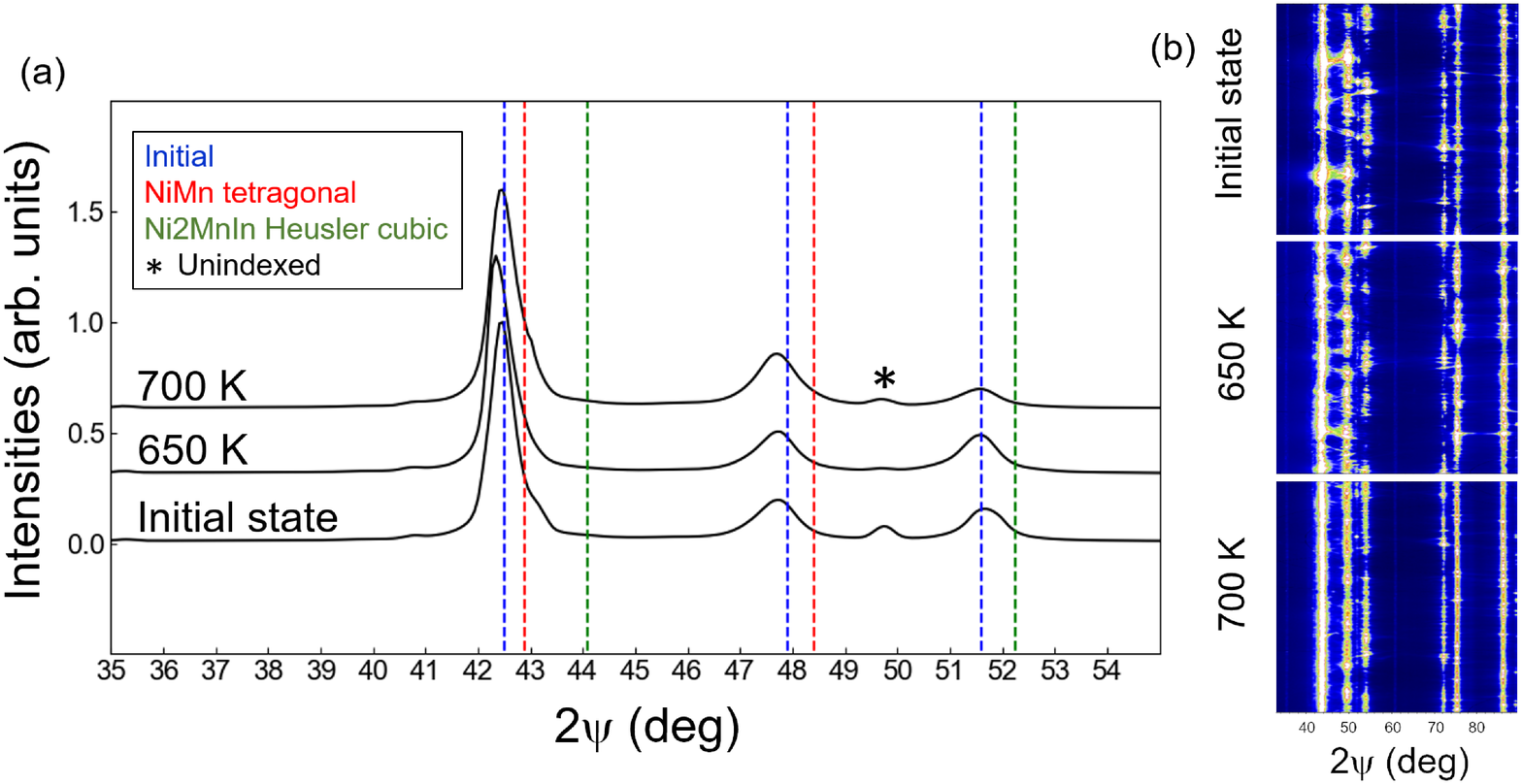}
\includegraphics[width=0.95\columnwidth]{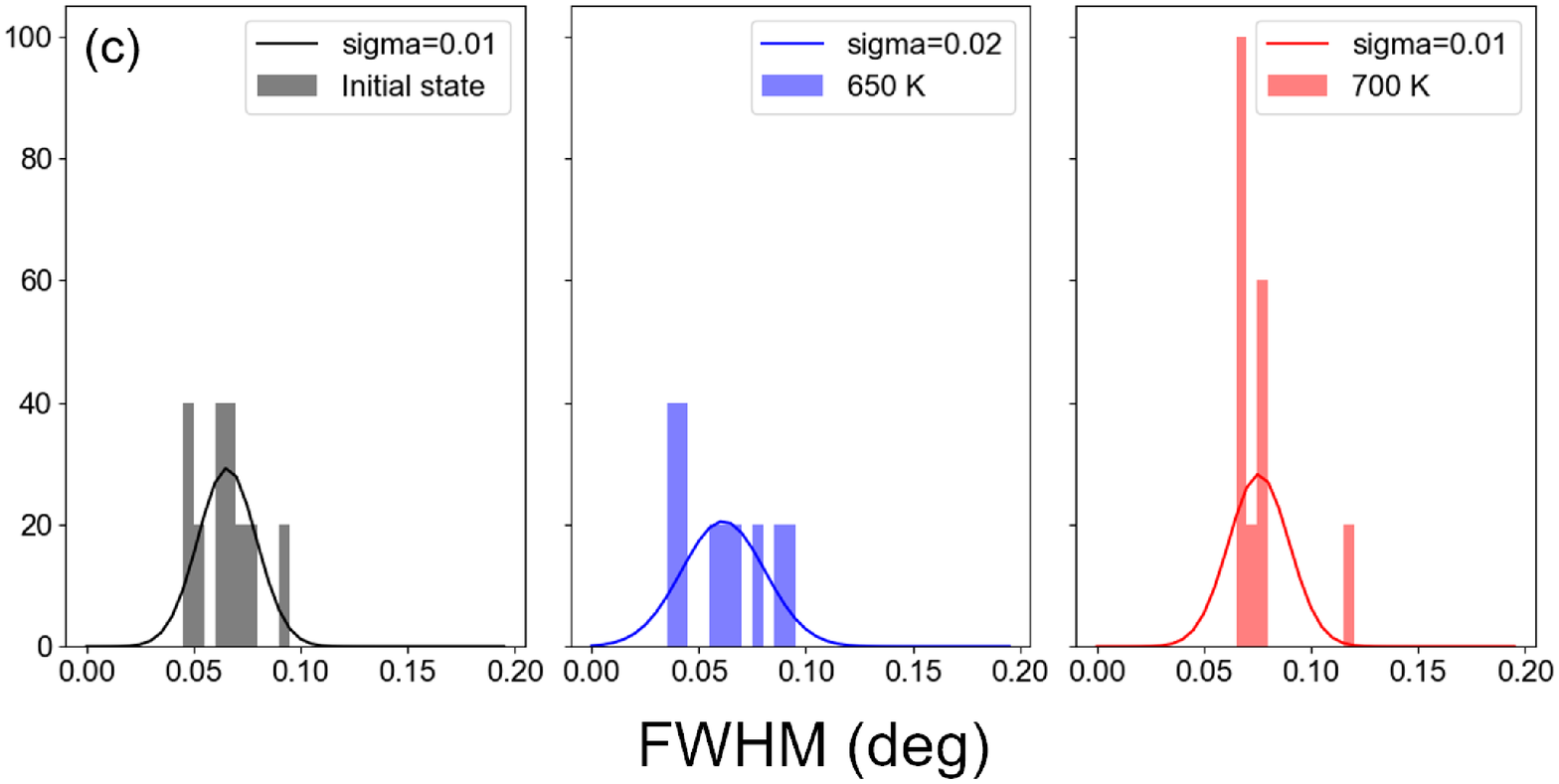}
\caption{(a)~Synchrotron x-ray diffraction data of field-annealed $\mathrm{Ni}_{50}\mathrm{Mn}_{45}\mathrm{In}_{5}$ (see inset). (b)~Two-dimensional diffraction data (transformed Debye-Scherrer rings) represented as a function of the scattering angle $\psi$ and the azimuthal angle $\theta$ (see text). (c)~Histograms of the FWHM of the most intense Bragg peaks. For the analysis and representation of the synchrotron data the Fit2D software has been used~\cite{fit2d}.}
\label{xrd}
\end{figure*}

The results of the magnetization measurements are shown in Fig.~\ref{MH1}. The initial state and the $650 \, \mathrm{K}$ samples both exhibit a paramagnetic-like behavior as the magnetization curves are nearly straight lines passing through the origin. This is consistent with the expected AF ground state of the initial-state $\mathrm{Ni}_{50}\mathrm{Mn}_{45}\mathrm{In}_{5}$ sample. By contrast, the $700 \, \mathrm{K}$ sample clearly shows, on top of a prevailing paramagnetic-like contribution, a ferromagnetic-like component as the magnetization curves exhibit hysteresis with a coercive field of $\sim 90 \, \mathrm{mT}$. This emerging feature is interpreted as the signature of the formation of dominantly ferromagnetic precipitates in an otherwise AF matrix. Additionally, both the $650 \, \mathrm{K}$ and $700 \, \mathrm{K}$ curves display a vertical shift (see inset in Fig.~\ref{MH1} for the $650 \, \mathrm{K}$ sample), depending on the orientation of the field with respect to the texture axis $\mathbf{c}$. This confirms the presence of a magnetic texture axis for both samples, in agreement with~\cite{cakir2016,cakir2017}.

\subsection{Synchrotron X-ray diffraction}

The synchtrotron x-ray diffraction results are summarized in Fig.~\ref{xrd}. The data follow the same trend as it has previously been reported in Fig.~2 of~\cite{cakir2016}. Note that the experimental scattering-angle values $\psi$ have been transformed into the corresponding values for $\mathrm{CuK}_\alpha$ radiation to ease the comparison with the data shown in~\cite{cakir2016}. The initial-state sample in Fig.~\ref{xrd}(a) (I4/mmm, $a = 3.79 \, \mathrm{\AA}$, $c = 7.00 \, \mathrm{\AA}$) decomposes with temperature treatment into a $\mathrm{Ni}\mathrm{Mn}$ tetragonal phase (I4/mmm, $a = 3.75 \, \mathrm{\AA}$, $c = 7.00 \, \mathrm{\AA}$). Also, one can see that the transformation seems to be more complete at $700 \, \mathrm{K}$ than at $650 \, \mathrm{K}$. However, in contrast to what has been reported in~\cite{cakir2016,dincklage2018} on powder samples, we cannot detect $\mathrm{Ni}_2\mathrm{MnIn}$ cubic Heusler nanoprecipitates (Fm--3m, $a \sim 5.8 \, \mathrm{\AA}$) in the diffraction pattern of the annealed samples. A possible explanation for this could be related to the presence of large grains and the nonideal particle statistics observed in the two-dimensional diffraction patterns, which might hinder the detection of small quantities of the nanoprecipitates. This becomes evident by inspection of Fig.~\ref{xrd}(b) and (c), which display for the three samples the evolution of the widths (FWHM) of the most intense diffraction peaks. For this analysis, each of the Debye-Scherrer rings has been transformed into a vertical line, containing data from the complete $360$ degrees. This way of presenting the Debye-Scherrer rings helps to identify crystallographic texture and poor particle statistics. The histograms in Fig.~\ref{xrd}(c) suggest that the particle statistics have improved for the $700 \, \mathrm{K}$ sample, i.e., the vertical lines are more homogenous, implying that the grain-size distribution has become less disperse after the heat treatment (in accordance with \cite{dincklage2018}). The presence of large crystals [large and intense spots in Fig.~\ref{xrd}(b)] is quite obvious for the initial-state and the $650 \, \mathrm{K}$ sample.

\subsection{SANS}

\subsubsection{Unpolarized SANS data}

\begin{figure*}[tb!]
\centering
\includegraphics[width=1.90\columnwidth]{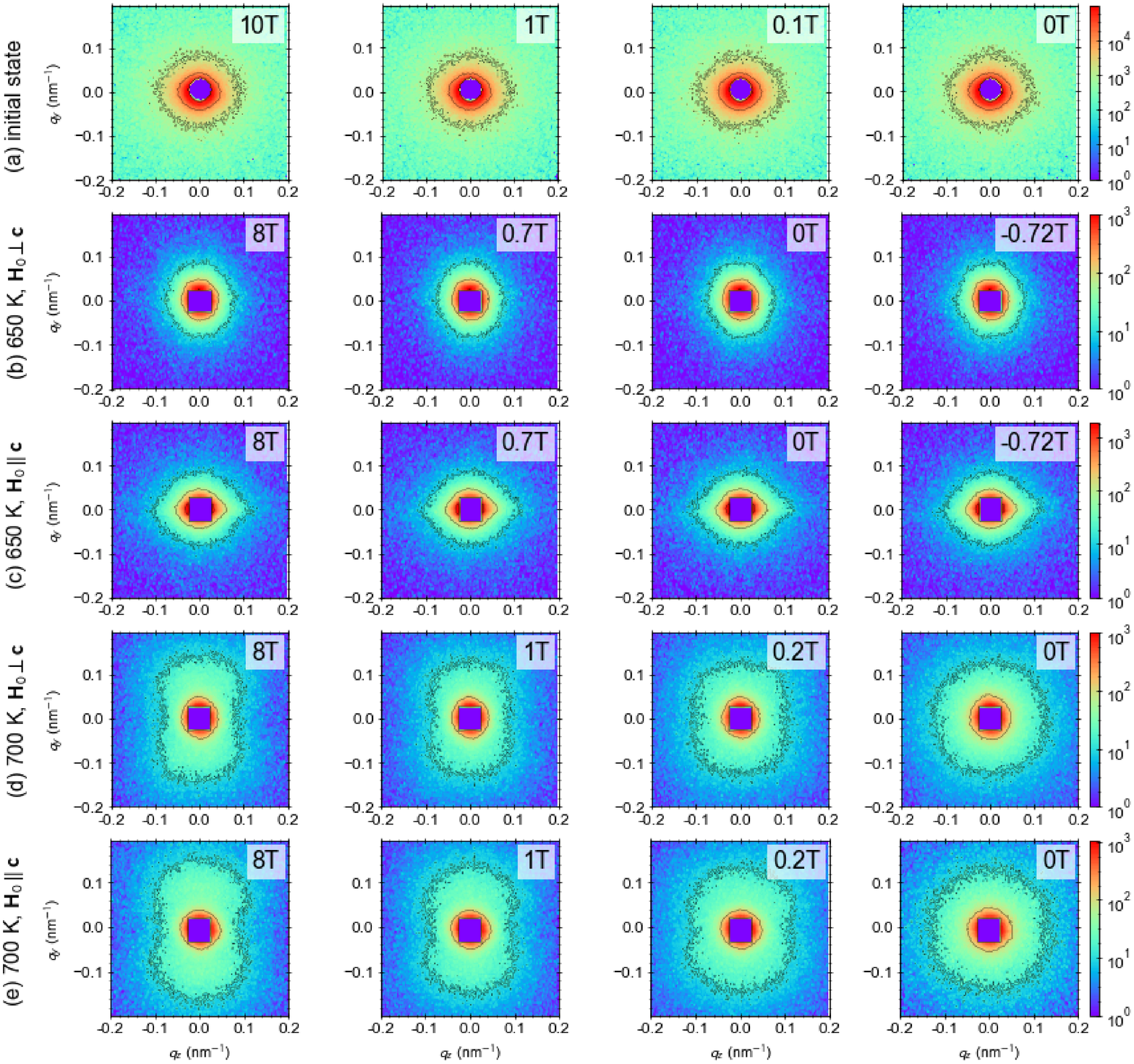}
\caption{Results for the two-dimensional unpolarized total (nuclear and magnetic) SANS cross section. Shown is $d \Sigma/ d \Omega$ at selected applied magnetic fields for the three studied samples (logarithmic color scale). $\mathbf{H}_0$ is horizontal in the plane of the detector ($\mathbf{H}_0 \perp \mathbf{k}_0$). (a)~initial state; (b)~$650 \, \mathrm{K}$, $\mathbf{H}_0 \perp \mathbf{c}$; (c)~$650 \, \mathrm{K}$, $\mathbf{H}_0 \parallel \mathbf{c}$; (d)~$700 \, \mathrm{K}$, $\mathbf{H}_0 \perp \mathbf{c}$; (e)~$700 \, \mathrm{K}$, $\mathbf{H}_0 \parallel \mathbf{c}$. Note that the $d \Sigma/ d \Omega$ scale is in arbitrary units for (a) and in absolute units ($\mathrm{cm}^{-1} \mathrm{sr}^{-1}$) for (b)$-$(e).}
\label{2D_SANS}
\end{figure*}

\begin{figure*}[tb!]
\centering
\includegraphics[width=1.90\columnwidth]{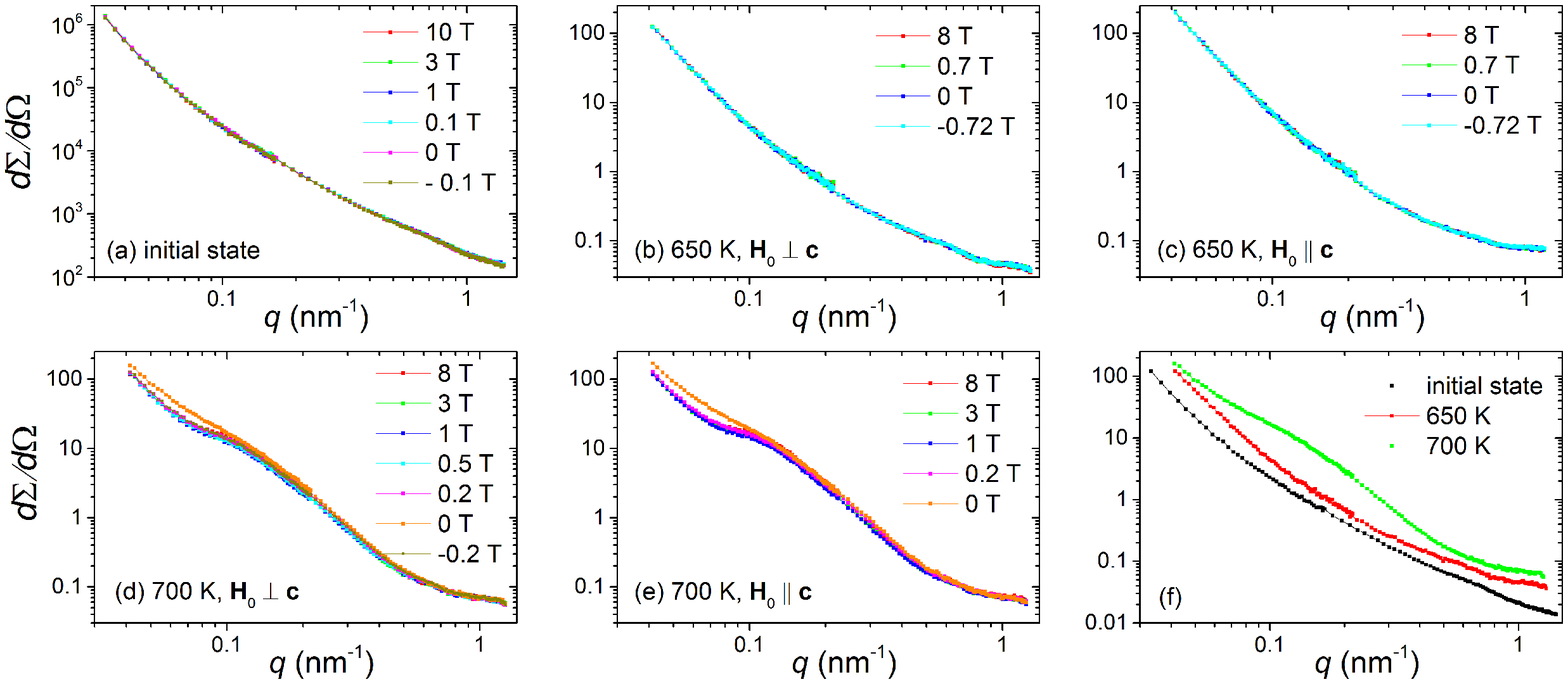}
\caption{Results for the azimuthally-averaged unpolarized SANS cross section $d \Sigma/ d \Omega$ at selected values of the applied magnetic field (see insets) (log-log scale). All $d \Sigma/ d \Omega$ are in absolute units ($\mathrm{cm}^{-1} \mathrm{sr}^{-1}$), except the data for the initial-state sample (a), which is available only in arbitrary units. (a)~initial-state sample; (b)~$650 \, \mathrm{K}$, $\mathbf{H}_0 \perp \mathbf{c}$; (c)~$650 \, \mathrm{K}$, $\mathbf{H}_0 \parallel \mathbf{c}$; (d)~$700 \, \mathrm{K}$, $\mathbf{H}_0 \perp \mathbf{c}$; (e)~$700 \, \mathrm{K}$, $\mathbf{H}_0 \parallel \mathbf{c}$; (f)~Comparison of $d \Sigma/ d \Omega$ for the initial state, $650 \, \mathrm{K}$ ($\mathbf{H}_0 \parallel \mathbf{c}$), and $700 \, \mathrm{K}$ ($\mathbf{H}_0 \parallel \mathbf{c}$) samples at zero field. Note that the data of the initial-state sample have been rescaled by a constant factor.}
\label{1D_SANS}
\end{figure*}

Figure~\ref{2D_SANS} depicts the two-dimensional unpolarized total (nuclear and magnetic) SANS cross sections $d \Sigma / d \Omega$ of the initial state, $650 \, \mathrm{K}$, and $700 \, \mathrm{K}$ samples at selected applied magnetic fields and for different orientations of the $\mathbf{c}$-axis relative to $\mathbf{H}_0$. The corresponding (over $2\pi$) azimuthally-averaged $d \Sigma / d \Omega$ are shown in Fig.~\ref{1D_SANS}. The initial state sample [Fig.~\ref{2D_SANS}(a)] exhibits an almost isotropic scattering pattern at all fields investigated. The azimuthally-averaged data sets [Fig.~\ref{1D_SANS}(a)] are field independent within the studied $(q, H_0)$-range. Since magnetic SANS is generally strongly field dependent~\cite{michels2014review}, this observation suggests a dominant nuclear (structural) scattering contribution to $d \Sigma / d \Omega$, in agreement with the AF state of the sample. Note that for an ideal defect-free AF the magnetization within a mesoscopic volume that is probed by SANS is zero, resulting in no magnetic SANS signal. Moreover, the shape of the $d \Sigma / d \Omega$ in Fig.~\ref{1D_SANS}(a) is distinctly different to the one typically found for particle form factors, i.e., a plateau at small momentum transfers followed by a Guinier and Porod regime cannot be discerned. Only a monotonous decay is visible, which indicates that there is significant scattering from large-scale structures.

The $650 \, \mathrm{K}$ sample mounted with its texture axis perpendicular to the field ($\mathbf{H}_0 \perp \mathbf{c}$) [Fig.~\ref{2D_SANS}(b)] displays a slightly anisotropic scattering pattern which is elongated along the direction normal to the field. The very weak angular anisotropy is also noticeable at the remanent state. When the sample is mounted with its texture axis parallel to the field ($\mathbf{H}_0 \parallel \mathbf{c}$) [Fig.~\ref{2D_SANS}(c)], then $d \Sigma / d \Omega$ has a maximum for directions parallel and antiparallel to the field. The azimuthally-averaged curves [Fig.~\ref{1D_SANS}(b) and (c)] reveal a field-independent $d \Sigma / d \Omega$, which (similar to the initial-state sample) points towards a negligible magnetic and a dominant nuclear SANS contribution. The origin for an anisotropic field-independent $d \Sigma / d \Omega$ could be related to the presence of shape-anisotropic structures, which might form as a consequence of the field-annealing process. The relatively sharp maxima in $d \Sigma / d \Omega$ for $\mathbf{H}_0 \parallel \mathbf{c}$ [Fig.~\ref{2D_SANS}(c)] resemble the so-called spike anisotropy, which was reported for a Nd-Fe-B permanent magnet~\cite{perigo2014}. Such sharp features in the magnetic SANS cross section can result from the magnetodipolar interaction due to the presence of $\mathbf{q} \neq 0$ Fourier modes of the magnetostatic field. However, the observations in Fig.~\ref{2D_SANS}(b) and (c) and in Fig.~\ref{1D_SANS}(b) and (c) that the $d \Sigma / d \Omega$ are field independent strongly suggest that the origin of the anisotropy is related to some structural feature. The present SANS data reveal no characteristic signature indicative of such presumably small precipitates, which might indicate that their size is too small and their volume fraction too low to give rise to a corresponding feature in the observed $q$-region. This is consistent with our synchrotron x-ray data analysis and with the results reported in~\cite{dincklage2018}, where precipitate sizes of the order of $3-5 \, \mathrm{nm}$ were found.

On application of a magnetic field, the $700 \, \mathrm{K}$ sample displays a strong $\sin^2\theta$-type anisotropy for both orientations of the texture axis $\mathbf{H}_0 \perp \mathbf{c}$ [Fig.~\ref{2D_SANS}(d)] and $\mathbf{H}_0 \parallel \mathbf{c}$ [Fig.~\ref{2D_SANS}(e)]. The scattering pattern in the remanent state is isotropic. These observations clearly reveal the presence of an effective ferromagnetic component in $d \Sigma / d \Omega$ [compare the term $|\widetilde{M}_z|^2 \sin^2\theta$ in Eq.~(\ref{sigmaPerpUnpol})] and are compatible with an annealing-induced formation of magnetic precipitates in an AF matrix. The azimuthally-averaged data [Fig.~\ref{1D_SANS}(d) and (e)] exhibit a broad shoulder at about $q \cong 0.1-0.15 \, \mathrm{nm}^{-1}$, which is indicative of the precipitation process, and a field-dependent SANS signal at the smallest $q$ demonstrating the existence of magnetic correlations on a length scale of at least several $10 \, \mathrm{nm}$. The direct comparison of the $d \Sigma / d \Omega$ of the different samples in the remanent state [Fig.~\ref{1D_SANS}(f)] highlights the effect of the heat treatment. The initial-state and $650 \, \mathrm{K}$ samples exhibit similar magnetization (see Fig.~\ref{MH1}) and scattering curves. When the sample is annealed at $700 \, \mathrm{K}$, then the $d \Sigma / d \Omega$ changes substantially: a broad hump at intermediate $q$ becomes visible, which we interpret as the signature of annealing-induced \textit{magnetic} precipitates. Note that for nonmagnetic precipitates in an AF matrix the magnetic scattering-length density contrast would vanish. Additionally, it is noted that all the $d \Sigma / d \Omega$ in Fig.~\ref{1D_SANS}(f) manifest significant scattering contributions at the smallest momentum transfers, originating from large-scale structures which cannot be resolved by our experimental setup ($q_{\mathrm{min}} \cong 0.04 \, \mathrm{nm}^{-1}$).

\subsubsection{Polarized SANS data}

\begin{figure}[b!]
\centering
\includegraphics[width=1.0\columnwidth]{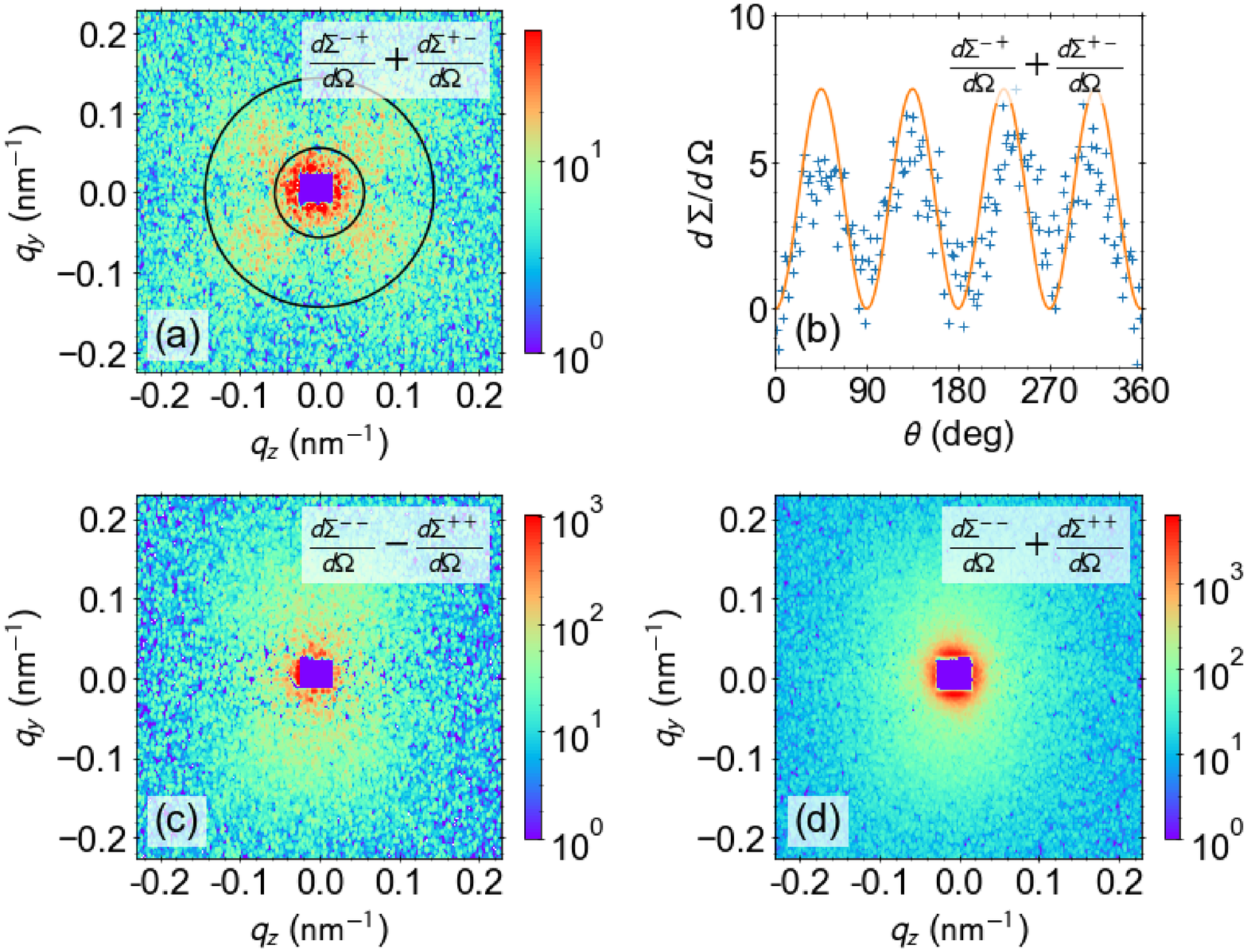}
\caption{Results for the two-dimensional spin-resolved SANS cross sections of the $700 \, \mathrm{K}$ sample at $\mu_0 H_0 = 1 \, \mathrm{T}$ (logarithmic color scale). (a)~Sum of the two spin-flip SANS cross sections $d \Sigma^{-+} / d \Omega + d \Sigma^{+-} / d \Omega$. (b)~Azimuthal $\theta$-dependence of $d \Sigma^{-+} / d \Omega + d \Sigma^{+-} / d \Omega$ for an average $q$-value from within the indicated black ring in (a). Solid line in (b): $d \Sigma^{-+} / d \Omega + d \Sigma^{+-} / d \Omega \propto \sin^2\theta \cos^2\theta$. (c)~Difference and (d)~sum of the two non-spin-flip SANS cross sections $d \Sigma^{--} / d \Omega$ and $d \Sigma^{++} / d \Omega$.}
\label{polaris_1Tno1}
\end{figure}

\begin{figure*}[tb]
\centering
\includegraphics[width=1.90\columnwidth]{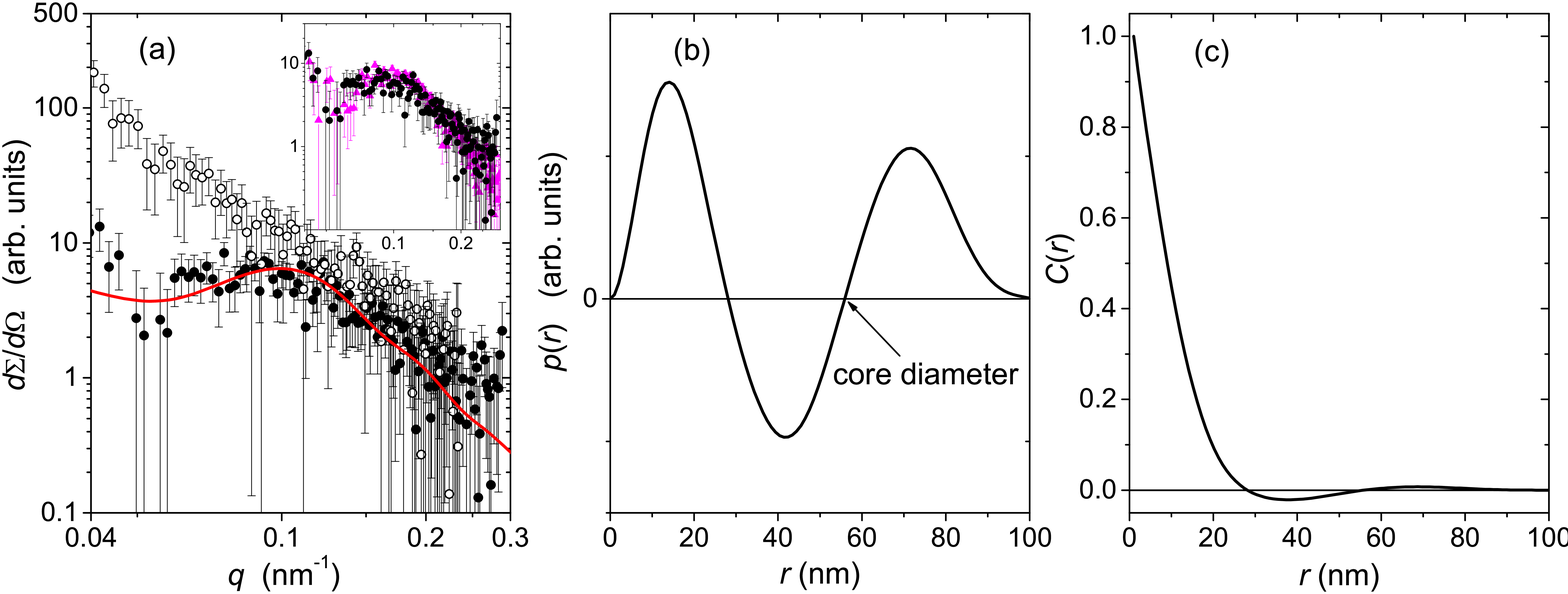}
\caption{(a)~Nuclear ($\circ$) and magnetic ($\bullet$) SANS cross sections of the $700 \, \mathrm{K}$ sample at $\mu_0 H_0 = 1 \, \mathrm{T}$ ($\mathbf{H}_0 \perp \mathbf{c}$) (log-log-scale). Red solid line:~Reconstructed magnetic scattering intensity based on the indirect Fourier transform from~(b). Inset in (a):~($\bullet$) Redrawn spin-flip cross section from~(a); (Magenta solid triangles):~Azimuthally-averaged data which results from the subtraction of the total unpolarized SANS at $8 \, \mathrm{T}$ from the total SANS at $1 \, \mathrm{T}$ (scaled to fit the spin-flip data). (b)~Distance distribution function $p(r) = r^2 C(r)$ of the spin-flip SANS cross section $d \Sigma^{+-} / d \Omega$. Arrow indicates the approximate particle core diameter. (c)~Normalized correlation function $C(r)$ of the spin-flip SANS cross section.}
\label{corrfunc}
\end{figure*}

Figure~\ref{polaris_1Tno1}(a) displays the sum of the two spin-flip SANS cross sections of the $700 \, \mathrm{K}$ sample at $\mu_0 H_0 = 1 \, \mathrm{T}$. Both spin-flip channels are equal within experimental uncertainty (data not shown), which implies the absence of chiral scattering contributions $\chi(\mathbf{q})$ [compare Eq.~(\ref{sigmaPerpPolaris2})]. The characteristic $\sin^2\theta \cos^2\theta$ anisotropy which becomes visible in $d \Sigma^{+-}/ d \Omega$ [Fig.~\ref{polaris_1Tno1}(b)] demonstrates that the spin-flip SANS at $1 \, \mathrm{T}$ is dominated by the longitudinal magnetization Fourier component $\widetilde{M}_z^2$. The spin-flip SANS is consistent with the unpolarized data and with the difference and the sum of the two non-spin-flip SANS cross sections, which are, respectively, displayed in Fig.~\ref{polaris_1Tno1}(c) and (d). Both combinations of cross sections manifest an anisotropic scattering pattern with maxima along the direction perpendicular to the applied field, in agreement with the nuclear-magnetic interference term $\propto \widetilde{N}\widetilde{M}_z \sin^2\theta$ in Eq.~(\ref{polarisdiff}) and the term $\propto |\widetilde{M}_z|^2 \sin^4\theta$ in Eq.~(\ref{polarissum}). This observation suggests that the effectively ferromagnetic particle phase which has formed during field annealing is to a good approximation spherically symmetric, such that the longitudinal magnetization Fourier component depends only on the magnitude of the scattering vector, i.e., $\widetilde{M}_z^2 = \widetilde{M}_z^2(q)$.

Figure~\ref{corrfunc}(a) displays the nuclear and the longitudinal magnetic SANS cross sections. The nuclear SANS was obtained by averaging the non-spin-flip SANS along the horizontal direction [$\theta = 0^{\circ}$ and $\theta = 180^{\circ}$], and the magnetic SANS results from a $2\pi$~azimuthal average of the spin-flip SANS $d \Sigma^{+-} / d \Omega$ at $1 \, \mathrm{T}$. As is explained in Sec.~\ref{polarissanscross}, such a separation is only possible by means of the polarization-analysis technique. The nuclear SANS signal exhibits a power-law behavior over the displayed $q$-range and shows no sign of particle scattering or a correlation peak (which might be expected in the case of a dense packing of nanosized particles), whereas the longitudinal magnetic SANS manifests a broad hump at around $q^{\star} = 0.1 \, \mathrm{nm}^{-1}$ with $2\pi / q^{\star} \cong 60 \, \mathrm{nm}$ followed by a small dip and a further increase at the smallest $q$. We have also subtracted the total unpolarized SANS cross section at a field of $8 \, \mathrm{T}$ from the corresponding data set at $1 \, \mathrm{T}$ (where the $d \Sigma^{+-} / d \Omega$ shown in Fig.~\ref{corrfunc}(a) was measured). The resulting unpolarized cross section [see inset in Fig.~\ref{corrfunc}(a)] is within error bars equal to $d \Sigma^{+-} / d \Omega$, in this way confirming that the observed functional dependency of the spin-flip scattering cross section (i.e., the broad hump centered at $\sim 0.1 \, \mathrm{nm}^{-1}$ and the dip at smaller $q$) is not an artifact related to the data reduction and the spin-leakage corrections involved in the POLARIS analysis.

In order to analyze this feature in the magnetic scattering in more detail, we have computed the corresponding correlation function $C(r)$, according to:
\begin{equation}
\label{1DcorrFunc}
C(r) = \int_0^{\infty} \frac{d \Sigma^{+-}}{d \Omega} j_0(qr) q^2 dq ,
\end{equation}
where $j_0(x) = \sin(x)/x$ denotes the zeroth-order spherical Bessel function. The correlation function $C(r)$ and the corresponding distance distribution function $p(r) = r^2 C(r)$ can be extracted by either a direct~\cite{bickapl2013,mettus2015,mettusprm2017,} or an indirect~\cite{hansen2000,svergun03,glatter2006,bender2017,hansenurl,bender2018prb} Fourier transform of $d \Sigma^{+-}/ d \Omega$.

Figure~\ref{corrfunc}(b) shows the computed $p(r)$. The profile of the distance distribution function disagrees with the expected nearly bell-shaped $p(r) = r^2 [1 - 3 r / (4R) + r^3 / (16 R^3)]$ of a homogeneous sphere with radius $R$~\cite{svergun03}. Rather, the computed oscillatory $p(r)$ is indicative of an \textit{inhomogeneous} core-shell-type particle, as has been shown by~\textcite{glatter1996}. To be specific, the oscillatory $p(r)$ of an isolated core-shell particle, exhibiting also negative values of $p(r)$, can only be reproduced if the scattering-length densities (slds) of the core, shell, and matrix either follow $\mathrm{sld}_{\mathrm{core}} < \mathrm{sld}_{\mathrm{matrix}} < \mathrm{sld}_{\mathrm{shell}}$ or $\mathrm{sld}_{\mathrm{shell}} < \mathrm{sld}_{\mathrm{matrix}} < \mathrm{sld}_{\mathrm{core}}$. In view of the fact that the matrix phase $\mathrm{Ni}_{50}\mathrm{Mn}_{50}$ in our sample is antiferromagnetic (zero net magnetization), corresponding to a zero magnetic sld of the matrix ($\mathrm{sld}^{\mathrm{mag}}_{\mathrm{matrix}} = 0$), the above combinations of slds require that either the magnetic sld of the core or the magnetic sld of the shell must be negative. Given that atomic magnetic scattering lengths are real-valued positive quantities, a negative magnetic sld cannot be realized. This implies that the $p(r)$ in Fig.~\ref{corrfunc}(b) cannot be related to an inhomogeneous particle exhibiting spatial variations in the \textit{magnitude} of the magnetization and, hence, of the magnetic sld.

\begin{figure}[tb]
\centering
\includegraphics[width=0.65\columnwidth]{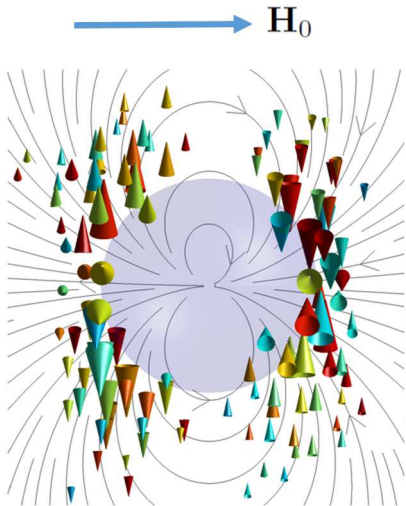}
\caption{Computed spin structure around a spherical pore in a ferromagnetic iron matrix (two-dimensional cross section out of a three-dimensional micromagnetic simulation; $D = 2R = 12 \, \mathrm{nm}$; $\mu_0 H_0 = 0.6 \, \mathrm{T}$). Shown is the magnetization component $\mathbf{M}_\perp(\mathbf{r})$ perpendicular to $\mathbf{H}_0$; thickness of arrows is proportional to the magnitude of $\mathbf{M}_\perp$. Solid grey lines: magnetodipolar field distribution. Image taken from~\textcite{erokhin2015}.}
\label{spinmis}
\end{figure}

However, as we argue in the following, the observed negative values of $p(r)$ can be explained by taking into account the vector character of the magnetization, i.e., by considering spatial variations in the \textit{orientation} of the magnetization, instead of variations in its magnitude. To support our statement we make reference to the micromagnetic simulation study by~\textcite{erokhin2015}, who have computed the correlation function for a distribution of spherical pores in an iron matrix. Figure~\ref{spinmis} shows the computed spin structure in the vicinity of a spherical pore at an applied magnetic field of $0.6 \, \mathrm{T}$. The distribution of iron spins decorates the dipolar stray fields of the pores and gives rise to a characteristic dipole-field-type spin texture. The ensuing correlation function $C(r)$ (Fig.~3 in~\cite{erokhin2015}) exhibits negative values, which was explained with the existence of ``anticorrelations'', i.e., the transversal magnetization component changes its sign along the direction of the applied field in this way giving rise to negative values of the correlation function at some field-dependent value. In Fig.~\ref{corrfunc}(c) the normalized $C(r) = p(r)/r^2$ of our sample is plotted at $\mu_0 H_0 = 1 \, \mathrm{T}$~\cite{pvonrvscvonr}. Qualitatively, it resembles the behavior of the $C(r)$ computed by~\textcite{erokhin2015}, suggesting that the data can indeed be explained by spatial nanometer-scale variations in the orientation of magnetic moments around the particles.

By analogy to the nuclear distance distribution function of a core-shell particle~\cite{glatter1996}, one can relate the second zero of $p(r)$ with the approximate particle (core) diameter. By comparison to Fig.~\ref{corrfunc}(b) this yields a value of $55 \, \mathrm{nm}$, while the size of the inhomogeneously magnetized region around the particle (corresponding to the ``shell thickness'' in the core-shell picture) is of the order of $20 \, \mathrm{nm}$. Both taken together, the homogeneously magnetized core and the nonuniformly magnetized shell region, represent what one may call the magnetic defect size. The value for the core diameter lies within the range of average precipitate sizes determined by the analysis of wide-angle x-ray diffraction data using the Scherrer formula~\cite{dincklage2018}.

The overall picture which emerges from these considerations is the following: the origin of the spin-flip SANS is related to the presence of essentially homogeneously magnetized particles in an antiferromagnetic matrix. The particles represent sources of perturbation, which give rise to canted spin moments in the surroundings of the particle-matrix interface, e.g., via inhomogeneous dipolar stray fields and/or strain fields. The average size of the particles is estimated at $55 \, \mathrm{nm}$, while the magnetically perturbed regions around the particles are of the order of a few $10 \, \mathrm{nm}$. 

\section{Conclusion and outlook}
\label{conclusion}

We have conducted unpolarized and polarized small-angle neutron scattering (SANS) experiments on magnetic-field-annealed bulk $\mathrm{Ni}_{50}\mathrm{Mn}_{45}\mathrm{In}_{5}$ Heusler-type alloys. For this material a core-shell-type ferromagnetic structure has recently been postulated based on the results of integral measurement techniques. High-energy synchrotron diffraction experiments could not detect the nanosized particles, most probably due to its low concentration and due to the presence of very large grains in the samples. However, spin-polarized SANS experiments have proved to be the ideal tool to study and characterize small and complex magnetic nanostructures. Our neutron data clearly reveals the precipitation of effectively ferromagnetic nanoparticles in a $\mathrm{Ni}_{50}\mathrm{Mn}_{45}\mathrm{In}_{5}$ sample when annealed in an applied magnetic field of $5 \, \mathrm{T}$ at a temperature of $700 \, \mathrm{K}$ during $12 \, \mathrm{h}$. Analysis of the spin-flip SANS cross section suggests that the nanoprecipitates are decorated by a region of nonuniformly magnetized spins. From the computation of the correlation function we estimate a value of $\sim 20 \, \mathrm{nm}$ for the magnetically inhomogeneous region surrounding the $\sim 55 \, \mathrm{nm}$-sized particles. Further data analysis using field-dependent spin-flip SANS and real-space techniques (e.g., Lorentz transmission electron microscopy) will help in identifying the nature (size, shape, structure) of the nanoprecipitates and the spin texture surrounding them.

\section*{Acknowledgements}

Giordano Benacchio, Philipp Bender, and Andreas Michels acknowledge financial support from the National Research Fund of Luxembourg (PRIDE MASSENA and CORE SANS4NCC grants). We acknowledge the MLZ, ILL, ANSTO, and the ESRF for provision of beamtime.

\appendix

\section{SANS cross sections}
\label{sans}

\begin{figure}[tb!]
\resizebox{1.0\columnwidth}{!}{\includegraphics{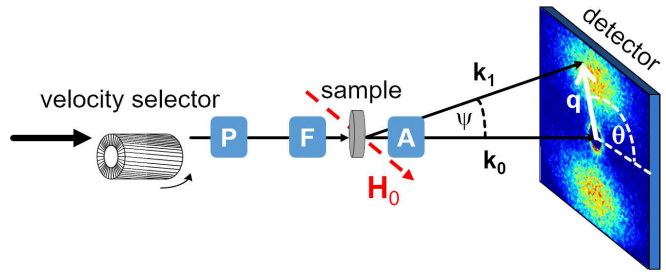}}
\caption{Schematic drawing of the SANS setup. The scattering vector $\mathbf{q}$ is defined as the difference between the wave vectors of the scattered and incident neutrons, i.e., $\mathbf{q} = \mathbf{k}_1 - \mathbf{k}_0$; its magnitude $q = |\mathbf{q}| = (4\pi/\lambda) \sin(\psi/2)$ depends on the mean wavelength $\lambda$ of the neutrons (selected by the velocity selector) and on the scattering angle $\psi$. The symbols ``P'', ``F'', and ``A'' denote, respectively, the polarizer, spin flipper, and analyzer, which are implemented in the POLARIS setup at D33. The applied-field direction $\mathbf{H}_0$ is parallel to the $\mathbf{e}_z$-direction of a Cartesian laboratory coordinate system and perpendicular to the incident neutron beam ($\mathbf{k}_0 \parallel \mathbf{e}_x \perp \mathbf{H}_0$). In the small-angle approximation, the component of $\mathbf{q}$ along $\mathbf{k}_0$ is neglected, i.e., $\mathbf{q} \cong \{0, q_y, q_z\} = q\{0, \sin\theta, \cos\theta\}$, where the angle $\theta$ specifies the orientation of $\mathbf{q}$ on the two-dimensional dectector.}
\label{sans_setup}
\end{figure}

\subsection{Unpolarized SANS}

For the scattering geometry where the applied magnetic field $\mathbf{H}_0$ (assumed to be parallel to the $\mathbf{e}_z$-direction of a Cartesian laboratory coordinate system) is perpendicular to the incident neutron beam ($\mathbf{H}_0 \perp \mathbf{k}_0$, compare Fig.~\ref{sans_setup}), the elastic unpolarized nuclear and magnetic SANS cross section $d \Sigma / d \Omega$ at momentum-transfer vector $\mathbf{q}$ can be written as~\cite{rmp2019}:
\begin{eqnarray}
\label{sigmaPerpUnpol}
\frac{d \Sigma}{d \Omega}(\mathbf{q}) = \frac{8 \pi^3}{V} b_H^2 \left( b_H^{-2} |\widetilde{N}|^2 + |\widetilde{M}_x|^2 + |\widetilde{M}_y|^2 \cos^2\theta \right. \nonumber \\ \left. + |\widetilde{M}_z|^2 \sin^2\theta - (\widetilde{M}_y \widetilde{M}_z^{\ast} + \widetilde{M}_y^{\ast} \widetilde{M}_z) \sin\theta \cos\theta \right) ,
\end{eqnarray}
where $V$ is the scattering volume, $b_H = 2.91 \times 10^{8} \, \mathrm{A^{-1} m^{-1}}$ is the atomic magnetic scattering length, and $\widetilde{N}(\mathbf{q})$ and $\widetilde{\mathbf{M}}(\mathbf{q}) = \{ \widetilde{M}_x(\mathbf{q}), \widetilde{M}_y(\mathbf{q}), \widetilde{M}_z(\mathbf{q}) \}$ denote, respectively, the Fourier transforms of the nuclear scattering-length density $N(\mathbf{r})$ and of the magnetization vector field $\mathbf{M}(\mathbf{r}) = \{ M_x(\mathbf{r}), M_y(\mathbf{r}), M_z(\mathbf{r}) \}$; $\theta$ represents the angle between $\mathbf{H}_0$ and $\mathbf{q}$ (see Fig.~\ref{sans_setup}); the asterisks ``$*$'' mark the complex-conjugated quantity and the atomic magnetic form factor (contained in the expression for $b_H$) is approximated to unity since we are dealing with forward scattering.

\subsection{Polarized SANS}
\label{polarissanscross}

Assuming perfect neutron optics and neglecting nuclear spin-incoherent SANS, the two non-spin-flip ($++$ and $--$) and the two spin-flip ($+-$ and $-+$) SANS cross sections of a bulk ferromagnet can, respectively, be expressed as ($\mathbf{H}_0 \perp \mathbf{k}_0$)~\cite{rmp2019}:
\begin{eqnarray}
\label{sigmaPerpPolaris1}
\frac{d \Sigma^{\pm\pm}}{d \Omega}(\mathbf{q}) = \frac{8 \pi^3}{V} b_H^2 \left( b_H^{-2} |\widetilde{N}|^2 + |\widetilde{M}_y|^2 \sin^2\theta \cos^2\theta \right. \nonumber \\ \left. + |\widetilde{M}_z|^2 \sin^4\theta - (\widetilde{M}_y \widetilde{M}_z^{\ast} + \widetilde{M}_y^{\ast} \widetilde{M}_z) \sin^3\theta \cos\theta \right. \nonumber \\ \left. \mp b_H^{-1} ( \widetilde{N}\widetilde{M}_z^{\ast} + \widetilde{N}^{\ast}\widetilde{M}_z) \sin^2\theta \right. \nonumber \\ \left. \pm b_H^{-1} ( \widetilde{N}\widetilde{M}_y^{\ast} + \widetilde{N}^{\ast}\widetilde{M}_y) \sin\theta \cos\theta \right) ,
\end{eqnarray}
\begin{eqnarray}
\label{sigmaPerpPolaris2}
\frac{d \Sigma^{\pm\mp}}{d \Omega}(\mathbf{q}) = \frac{8 \pi^3}{V} b_H^2 \left( |\widetilde{M}_x|^2 + |\widetilde{M}_y|^2\cos^4\theta \right. \nonumber \\ \left. + |\widetilde{M}_z|^2 \sin^2\theta \cos^2\theta \right. \nonumber \\ \left. - (\widetilde{M}_y \widetilde{M}_z^{\ast} + \widetilde{M}_y^{\ast} \widetilde{M}_z) \sin\theta \cos^3\theta \mp i \chi \right).
\end{eqnarray}

The first superscript (e.g., $+$) that is attached to $d \Sigma / d \Omega$ in Eqs.~(\ref{sigmaPerpPolaris1}) and (\ref{sigmaPerpPolaris2}) refers to the spin state of the incident neutrons, whereas the second one (e.g., $-$) specifies the spin state of the scattered neutrons. The direction of $\mathbf{H}_0 \parallel \mathbf{e}_z$ specifies the quantization axis for the neutron spins. The two spin-flip channels [Eq.~(\ref{sigmaPerpPolaris2})] depend on the initial neutron polarization only via the chiral function $\chi(\mathbf{q})$. The polarization-dependent terms $\propto \widetilde{N} \widetilde{M}_y \sin\theta \cos\theta$ in the two non-spin-flip cross sections [Eq.~(\ref{sigmaPerpPolaris1})] average out for statistically-isotropic polycrystalline magnetic materials, since there are no correlations between spatial variations in the nuclear density and in the transversal magnetization components (note that $\langle M_y \rangle = V^{-1} \int_V M_y(\mathbf{r}) dV = 0$ for such a material). The difference between $\frac{d \Sigma^{--}}{d \Omega}$ and $\frac{d \Sigma^{++}}{d \Omega}$ then reads:
\begin{eqnarray}
\label{polarisdiff}
\frac{d \Sigma^{--}}{d \Omega} - \frac{d \Sigma^{++}}{d \Omega} \propto \left( \widetilde{N}\widetilde{M}_z^{\ast} + \widetilde{N}^{\ast}\widetilde{M}_z \right) \sin^2\theta ,
\end{eqnarray}
while their sum is given by
\begin{eqnarray}
\label{polarissum}
\frac{d \Sigma^{--}}{d \Omega} + \frac{d \Sigma^{++}}{d \Omega} \propto \left( |\widetilde{N}|^2 + |\widetilde{M}_y|^2 \sin^2\theta \cos^2\theta \right. \nonumber \\ \left. + |\widetilde{M}_z|^2 \sin^4\theta - (\widetilde{M}_y \widetilde{M}_z^{\ast} + \widetilde{M}_y^{\ast} \widetilde{M}_z) \sin^3\theta \cos\theta \right) .
\end{eqnarray}

Inspection of the two non-spin-flip SANS cross sections [Eq.~(\ref{sigmaPerpPolaris1})] shows that their evaluation at angles $\theta = 0^{\circ}$ and $\theta = 180^{\circ}$ yields the nuclear SANS cross section $d \Sigma_{\mathrm{nuc}} / d \Omega = \frac{8 \pi^3}{V} |\widetilde{N}|^2$. This is a particular strength of the polarization-analysis technique~\cite{michels2010epjb,michels2012prb2}. Performing the corresponding evaluation of the unpolarized SANS cross section [Eq.~(\ref{sigmaPerpUnpol})] yields the nuclear SANS only if the magnetic microstructure is completely saturated, i.e., when $\mathbf{M} = \{ 0, 0, M_z = M_s(\mathbf{r}) \}$. For only partially saturated samples, the average of the total unpolarized SANS along the horizontal direction contains ``contaminations'' due to misaligned spins [cf.\ the term $|\widetilde{M}_y|^2 \cos^2\theta$ in Eq.~(\ref{sigmaPerpUnpol})]. Moreover, from the measurement of the spin-flip SANS cross section we obtain the purely magnetic scattering (no nuclear coherent SANS). This circumstance is of central importance for reaching the conclusions on the magnetic microstructure of the present $\mathrm{Ni}_{50}\mathrm{Mn}_{45}\mathrm{In}_{5}$ Heusler alloy.

Polarized neutrons are very useful in the study of magnetic materials, for instance, for separating weak magnetic signals from strong nuclear scattering~\cite{keller2000}. This is because experiments with a polarized incident neutron beam allow for the measurement of scattering terms that depend only linearly on the nuclear and magnetic scattering amplitudes [compare Eq.~(\ref{sigmaPerpPolaris1})], instead of the usual quadratic dependence of the total unpolarized $d \Sigma / d \Omega$ on these terms [compare Eq.~(\ref{sigmaPerpUnpol})]. In fact, the difference between the non-spin-flip measurements, $d \Sigma^{--} / d \Omega - d \Sigma^{++} / d \Omega$, depends on the nuclear-magnetic interference term $\widetilde{N} \widetilde{M}_z \sin^2\theta$ [Eq.~(\ref{polarisdiff})], while $d \Sigma^{+-} / d \Omega - d \Sigma^{-+} / d \Omega$ yields the chiral function $\chi(\mathbf{q})$. Furthermore, the two spin-flip SANS cross sections $d \Sigma^{\pm\mp} / d\Omega$ are free of nuclear coherent scattering and depend only on the magnetic scattering amplitudes $\widetilde{M}_{x,y,z}$ [compare Eq.~(\ref{sigmaPerpPolaris2})]. Since the cross section is a scalar quantity and the incident polarization $\mathbf{P}_0$ is an axial vector, it is clear that the system under study must itself contain an axial vector in order that the cross section depends on $\mathbf{P}_0$~\cite{maleyev2002}. Examples for such ``built-in'' vectors are related to the interaction of a polycrystalline sample with an external magnetic field (inducing an average magnetization directed along the applied field), the existence of a spontaneous magnetization in a ferromagnetic single crystal, the antisymmetric Dzyaloshinskii-Moriya interaction, mechanical (torsional) deformation, or the presence of spin spirals~\cite{maleyev2002}. If on the other hand there is no preferred axis in the system, then the cross section is independent of $\mathbf{P}_0$; examples include a collection of randomly oriented nuclear (electronic) spins, which describe the general case of nuclear (paramagnetic) scattering at not too low temperatures and large applied fields, or a multi-domain ferromagnet with a random distribution of the domains. In the present study, the initial-state and the $650 \, \mathrm{K}$ sample did not show a polarization dependence of the SANS cross section, which is consistent with the paramagnetic-like magnetization response of both samples (compare Fig.~\ref{MH1}).

\bibliographystyle{apsrev4-2}

%

\end{document}